\documentclass[nofootinbib,tightenlines,nobibnotes, aps,
amssymb,amsmath,prb]{revtex4}
\usepackage{graphicx}% Include figure files 
\usepackage{epsfig}% Include figure files

\begin{document}
\preprint{APS/123-QED}
%draft
%\twocolumn[\hsize\textwidth\columnwidth\hsize\csname @twocolumnfalse\endcsname
\title{Quenched Averages for self-avoiding  walks and polygons on 
deterministic fractals} 
\author{Sumedha} 
\altaffiliation[Present address:]{Laboratoire de
Physique Th$\acute{e}$orique et Mod$\acute{e}$les Statistiques,
Universit$\acute{e}$ Paris-sud,  F-91405, France}
%\affiliation{  
%Laboratoire de
%Physique Th$\acute{e}$orique et Mod$\acute{e}$les Statistiques,
%Universit$\acute{e}$ Paris-sud,  F-91405, France 
%}

\author{Deepak Dhar}
\affiliation{% 
Department of Theoretical Physics, Tata Institute of
Fundamental Research, Homi Bhabha Road, Colaba, 
Mumbai 400005, India 
}%
\date{\today}
                                                                                
\begin{abstract}

We study rooted self avoiding polygons and self avoiding walks on
deterministic fractal lattices of finite ramification index. Different sites
on such lattices are not equivalent, and the number of rooted open walks
$W_n(S)$, and rooted self-avoiding polygons $P_n(S)$ of $n$ steps depend on
the root $S$.  We use exact recursion equations on the fractal to determine
the generating functions for $P_n(S)$, and $W_n(S)$ for an arbitrary point
$S$ on the lattice. These are used to compute the averages $< P_n(S)>,
<W_n(S)>, <\log P_n(S)>$ and $<\log W_n(S)>$ over different positions of
$S$. We find that the connectivity constant $\mu$, and the radius of
gyration exponent $\nu$ are the same for the annealed and quenched
averages. However, $<\log P_n(S)> \simeq n \log \mu + (\alpha_q -2 ) \log
n$, and $<\log W_n(S)> \simeq n \log \mu + (\gamma_q -1) \log n$, where the
exponents $\alpha_q$ and $\gamma_q$ take values different from the annealed
case. These are expressed as the Lyapunov exponents of random product of
finite-dimensional matrices.  For the $3$-simplex lattice, our numerical
estimation gives $ \alpha_q \simeq 0.72837 \pm 0.00001$; and $\gamma_q \simeq
1.37501 \pm 0.00003$, to be compared with the annealed values $\alpha_a =
0.73421$ and $\gamma_a = 1.37522$.

\end{abstract}
                                                                                
%\pacs{74.20.Mn 74.30.+h 74.20.-z 71.55.Jv} 
%\vskip2pc 
%\keywords{ disorder average, fractals, self-avoiding walk}
                                                                                
\maketitle

\section{Introduction}

Understanding the behavior of linear polymers in random media has been an
important problem in statistical physics, both for reasons of theoretical
interest, and applications. Calculation of quenched averages over the
disorder  is a very hard problem both analytically and computationally
\cite{barat,chakrabarti}. There have been many speculations and controversies
regarding critical behavior of self-avoiding walks (SAWs) in random media
(especially at the percolation threshold).  Exact calculation of quenched
averages has not been possible so far for any nontrivial case, and
simulations are not easy and often give contradicting results for this
problem. In this context, it seems useful to construct a toy model, where
one can explicitly calculate the quenched and annealed averages, and see
their difference. This is what we shall do in this paper, for the problem of
linear polymers on a deterministic fractal.

It is straight forward to calculate annealed averages of self-avoiding walks
on deterministic fractals using real-space renormalization techniques
\cite{dd78, rammal}. One can explicitly write down a
closed set of exact renormalization equations in a finite number of
variables, so long as the fractal has a finite ramification index. Then, 
the eigenvalues of the linearized recursion equations near the fixed point 
of the renormalization transformation determine the critical exponents of 
the problem. In fact, a good deal of understanding of the complex behavior 
of polymers with additional interactions, e.g. self-interaction, or with a 
wall, or with other polymers, has been obtained by studying the 
corresponding analytically tractable problem on  fractals \cite{ddysrev}.

It seems reasonable that the study of effect of inhomogeneities of the
substrate would also be more tractable on fractal lattices.  This is
specially promising, as one does not need to introduce disorder in the
problem from outside.  The fractal lattices do not have translational
symmetry, and hence a polymer living on a fractal lattice necessarily sees
an inhomogeneous environment.  Some regions of the lattice are better
connected than others, and the local free density of the polymer per 
monomer in these regions would be lower than other parts. We want to 
understand the effect of presence of such regions on the large-scale 
structure and properties of the polymer. The main difference from the 
usual polymer in disordered medium problem to the case we study here 
is that the favorable and unfavorable regions are not randomly distributed 
over the lattice, but have a predetermined regular structure in the 
case of deterministic fractals. In this context, the annealed 
averages, which means averaging the partition function of the 
polymer over different positions, are appropriate in cases where 
the polymer can move freely over different parts of the lattice. 
The quenched average is averaging {\it the logarithm} of the partition 
function of the polymer over different positions, and would be appropriate 
where this freedom is not present.

 The annealed average for linear and branched polymers have been calculated
exactly for many different fractal lattices \cite{ddysrev}. But to our
knowledge the quenched averages have not been calculated so far on any
fractal lattice. In this paper, we use the recursive structure of fractals
to calculate quenched averages for linear polymers on deterministic fractals. 
We find that the connectivity constant $\mu$, and the radius of gyration 
exponent $\nu$ are the same for the annealed and quenched averages. The critical
exponents for the quenched case can be expressed as the Lyapunov exponents for
random product of finite-dimensional matrices. These can be estimated
numerically efficiently by Monte Carlo methods, which we do for the
illustrative case of $3$-simplex fractal.

The rest of the paper is organized as follows: In section 2 we define the
$3$-simplex lattice and specify the scheme we use to label the sites of the
lattice. We also define the annealed and quenched averages precisely, and
the generating functions for different quantities of interest. In section 3
we work out the recursion equations for generating functions of 
self-avoiding polygons(SAPs) and SAWs on $3$-simplex. In section 4 we 
derive rigorous bounds on the number of rooted SAPs and SAWs, and prove 
that the connectivity constant $\mu$, and the size exponent $\nu$
are the same for the quenched and annealed averages. In section 5, we study 
the variation of the number of SAPs and SAWs with the position of the
root on the $3$-simplex lattice numerically. In section 6, we determine
numerical values of the critical exponents in the quenched case by Monte
Carlo determination of Lyapunov exponents for random products of matrices.

\section{Preliminaries and definitions}

We will illustrate the general technique by working out explicitly the 
simple case of SAWs and SAPs on the $3$-simplex fractal. The treatment 
is easily generalized to other recursively defined fractals of finite 
ramification index.

\begin{figure}
\epsfxsize=.7\hsize\centerline{\epsfbox{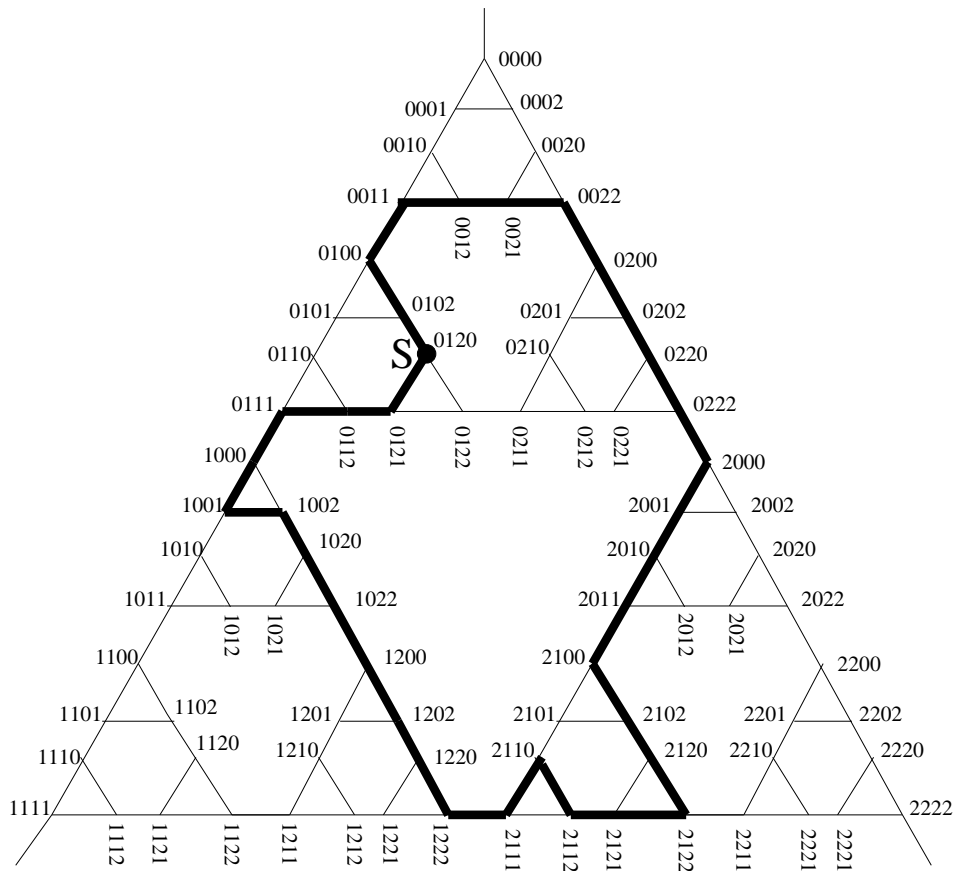}}
\caption{The figure shows a $3$ simplex graph of order $5$, which is formed 
by joining $3$ fourth order graphs. In general the $r$th order $3$-simplex 
is formed by joining $3$ graphs of $(r-1)$th order such that there is one
dangling bond left in each subgraph.The thicker line shows a polygon of size
$35$ which passes through a particular site $S$.
\label{simplex1}}
\end{figure}

The $3$-simplex graph is defined recursively as follows [Fig.\ref{simplex1}]: 
the graph of the first order triangle is a single vertex with $3$ 
bonds. The $(r+1)$th order triangle is formed by joining graphs of three 
$r$-th order triangle by connecting a dangling bond of each to a dangling bond 
of the other $r$th order subgraphs. There is one dangling bond left in each 
graph and $3$ bonds altogether. In general, the $r$th order graph will have 
$3^{r-1}$ vertices and $(3^r-3)/2$ internal bonds, and $3$ boundary bonds. 
Clearly, the fractal dimension of this lattice is $\log 3/\log 2$.

We use a ternary base single integer to label different sites of an $r$-th 
order triangle. The labeling is explained in Fig. \ref{simplex1}. A point 
on the $r$-th order triangle is labelled by a string of $(r-1)$ 
characters, e.g. $0122201 \ldots$. Each character takes one of three 
values $0,1$ or $2$. The leftmost character specifies in which of the three
sub-triangles the point lies ($0$, $1$ and $2$ for the top, left and right
sub-triangle respectively).  The next character specifies placement in the
$(r-1)$-th order sub-triangle, and so on. On an infinite lattice,
specification of $S$ requires an infinitely long string. In discussing the
local neighborhood of  a site, we only need to know the last few digits of
$S$. We will denote by $[S]_r$  the substring consisting of the last $r$
characters of  the integer label  of $S$, and by  $s_{r}$  the rth digit in
the string  $S$  counted  from the right. As an example, for the string $S=
21....0112011$, $[S]_2 =  11$, and $s_3 = 0$. Also, in an obvious notation,
$[S]_r = s_r [S]_{r-1}$.

The rooted polygons are polygons that pass through a given site (see Fig.
\ref{simplex1}) called the root. Let $P_n(S)$ be the number of rooted
polygons of  perimeter $n$ corresponding to root  $S$.  The generating
function  for rooted polygons for root $S$ is defined as

\begin{equation}
P(x;S) = \sum_{n=3}^{\infty} P_n(S) x^n \label{eq1}
\end{equation}

Similarly, we define $W_n(S)$ as the number of open walks of length $n$
whose one end-point  is $S$, and the corresponding generating function
$W(x;S)$ by

\begin{equation}
W(x;S) = \sum_{n=1}^{\infty} W_n(S) x^n   \label{eq2}
\end{equation}

For a given value of $n$, the values of $P_n(S)$ and $W_n(S)$ depend only on
the last few digits in the ternary integer labeling $S$. For example, it is
easy to check that if the last two digits of $S$ are unequal, i.e. if $s_1 \neq
s_2$ we have
\begin{equation}
P(x;S) = x^3 + x^6 +3 x^7 + 3 x^8 + x^9 + {\cal O}(x^{12}).
\end{equation}
But if $s_1 = s_2$
\begin{equation}
P(x;S) = x^3 +   x^7 + 2 x^8 + x^9 + {\cal O}(x^{12})
\end{equation}

In general, for any SAP configuration consisting of perimeter $n$, with $ n
< 3 \times 2^{r} $, with $r$ an integer, one can find an $(r+1)$-th order 
triangle graph, such that the polymer lies completely inside it. Then, the 
numbers $P_n(S)$ and $W_n(S)$ depend only the relative position of $S$ within 
this triangle, and hence only on $[S]_r$. This then allows us to define averages
of functions of $P_n(S)$ and $W_n(S)$. One assumes that $S$ is equally
likely to be any one of the $3^r$ sites in the $(r+1)$-th order
triangle. So, for example, for $n=7$, we have $r=2$, and $P_7(S)$ depends
only on $[S]_2$. Out of $9$ possibilities for $[S]_2$, $3$ have $s_1 =s_2$,
and $6$ cases have $s_1 \neq s_2$.  Thus, if $S$ is chosen at random, using 
Eq. (3-4), we have $Prob[P_7(S)=3] = 2/3$, and $Prob[P_7(S)=1] = 1/3$. We 
shall use angular brackets $\langle \rangle$ to denote averaging over 
different positions of the root $S$. This gives $\langle P_7 \rangle = 7/3$, 
and $\langle \log P_7 \rangle = \frac{2}{3} \log 3$. Other averages can be
calculated similarly.

We shall call the values $\langle P_n(S) \rangle$, and  $\langle
W_n(S) \rangle$ as the annealed averages of  $P_n(S)$ and
$W_n(S)$, and define their generating functions
\begin{equation}
\bar{P}(x) = \sum_{n=3}^{\infty} \langle{P}_n(S)\rangle  x^n  \label{eq3}
\end{equation}
\begin{equation}
\bar{W}(x) = \sum_{n=1}^{\infty} \langle {W}_n (S)\rangle  x^n   
\label{eq4}
\end{equation}

We define the growth constant $\mu_a$, and the annealed exponents $\alpha_a$
and $\gamma_a$ in terms of the behavior of $\langle{P}_n(S) \rangle$ and 
$\langle {W}_n(S) \rangle$ for
large $n$:
\begin{eqnarray}
\log {\langle P_n (S)\rangle} &=& n \log \mu_a + (\alpha_a -2) \log n + 
{\cal 
O}(1)
\label{eq7}\\ 
\log {\langle W_n(S)\rangle} &=& n \log \mu_a + (\gamma_a-1) \log n + 
{\cal 
O}(1)
\label{eq8} 
\end{eqnarray}
It was shown in ref.3 that  $\mu_a \approx 1.61803$, $\alpha_a \approx 
0.73421$ and $\gamma_a \approx 1.37522$. 

For the quenched averages, the exponents $\alpha_q$ and $\gamma_q$ are 
defined by the condition that for large $n$
\begin{eqnarray}
\langle \log P_n(S) \rangle &=& n \log \mu_q + (\alpha_q -2) \log n  +{\cal
O}(1)
\label{eq9a}\\
\langle \log W_n(S) \rangle &=& n \log \mu_q + (\gamma_q-1) \log n  +{\cal
O}(1)
\label{eq10a}
\end{eqnarray}

We define the order of a polygon as the order of the smallest triangular
subgraph that contains all the sites occupied by the polygon. For defining
the size exponent $\nu_a$ and $\nu_q$, it is sufficient to adopt the simple
definition that the diameter of a polygon is $2^r$ if its order is $r$. One
can then define the mean diameter of all polygons of perimeter $n$ rooted at
a given site $S$, as the average diameter, with all such polygons given
equal weight. We define the quenched average mean diameter as the average
over different positions of $S$ of the average diameter of polygons of
perimeter $n$ rooted at $S$. The size-exponent $\nu_q$ for quenched
averages is defined by the condition that the quenched average diameter
varies as $n^{\nu_q}$ for large $n$.

To define the annealed average for the diameter of SAP of perimeter $n$, 
we assign equal weight to all such loops within an $s$th order triangle with
$2^s > n$, and calculate the average diameter. It is easy to see that the
answer does not depend on $s$. Again, we define the annealed size exponent
$\nu_a$ by the condition that the annealed average diameter varies as
$n^{\nu_a}$ for large $n$.

The exponents for  open walks can be defined similarly. We shall argue that
the size exponents for open walks and polygons are the same,  and further
that $\mu_q = \mu_a$, and $\nu_q = \nu_a$, and hence simply write $\mu$ and
$\nu$ without any subscript, if the distinction is unnecessary.

Note that we have to first  average over different positions of the root
for a fixed $n$, and then let $n$ tend to infinity to define
$\nu_q$. If we take the large $n$ limit first, for a fixed position of 
the root, then even the convergence of large $n$ limit of 
$\log[P_n(S) \mu^{-n}]/\log n$ and $\log[W_n(S) \mu^{-n}]/\log n$ is not obvious 
due to the irregular variation of $\log P_n(S)$ and $\log W_n(S)$ with $n$. 
The amount of fluctuations in different averages of observables over 
different positions  of the root will be discussed in section 5.
 
\section{The renormalization equations}

In this section, we  briefly recapitulate the  renormalization  scheme for
calculating the annealed averages used in ref. 3, and then adapt it for
calculating properties of rooted walks.

\begin{figure}
\epsfxsize=.7\hsize\centerline{\epsfbox{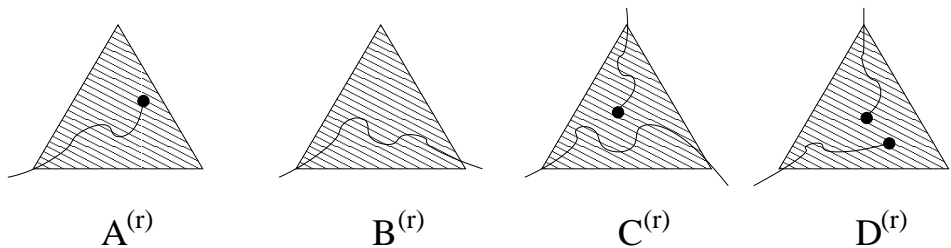}}
\caption{Different restricted partition functions for the $r$th order
triangle. Internal vertices inside the $r$th order triangle are not  shown.}
\label{3simplex2}
\end{figure}

Consider one $r$-th order triangular subgraph of the infinite order graph.
It is connected to the rest of the lattice by only three bonds. Our aim is
to sum over different configurations of the SAW that lie within the
subgraph, with  a weight $x$ for each step of the walk. These 
configurations can be
divided into  four classes, as shown in Fig. \ref{3simplex2}, and  we define
four restricted  partition functions $A^{(r)}, B^{(r)}, C^{(r)}$ and
$D^{(r)}$  corresponding to these four classes.

Here $A^{(r)}$ is the sum over all configurations of the walk within the
$r$-th order triangle, that enters the triangle from a specified corner, and
with one endpoint inside the triangle. $B^{(r)}$ is the sum over all
configurations of walk within the triangle that enters and leaves the
triangle from specified corner vertices. $C^{(r)}$ and $D^{(r)}$ are defined
similarly (Fig. \ref{3simplex2}). For any given value of $r$, $B^{(r)}, 
D^{(r)}, A^{(r)} \sqrt{x}, C^{(r)} \sqrt{x}$ are finite degree polynomials 
in $x$ with non-negative integer coefficients. It is easy to see that the 
starting values of these variables are
\begin{equation} 
A^{(1)} = \sqrt{x}, ~B^{(1)} = x, ~C^{(1)} =~ D^{(1)} =0.
\end{equation}

We note the generating function for $\bar{P}(x)$ is simply related to the
generating function of unrooted polygons $P_{noroot}(x)$ by the relation
$\bar{P}(x) = x \frac{d}{dx} P_{noroot}(x)$. The sum over all unrooted
polygons of order $(r+1)$ within an $(r+1)$-th order triangle is ${B^{(r)}}^3$, 
and number of sites in the $(r+1)$-th order triangle is $3^{r}$, hence we 
get

\begin{equation}
P_{noroot}(x) = \sum_{r=1}^{\infty} 3^{-r}  {B^{(r)}}^3.
\label{eq11}
\end{equation}
here we have suppressed the $x$ dependence of $B$. In rest of the paper,
we will suppress the $x$-dependence of variables $A, B, C, D$, to simplify 
notation, whenever the meaning is clear from the context.
 
The sum over unrooted open walks can be expressed similarly
\begin{equation}
W_{noroot}(x) =  \sum_{r=1}^{\infty} 3^{-r} [ 3 {A^{(r)}}^2 + 3 B^{(r)}
{A^{(r)}}^2 + 3 {B^{(r)}}^2 D^{(r)} ].
\label{eq12}
\end{equation}
   
\begin{figure}
\epsfxsize=.7\hsize\centerline{\epsfbox{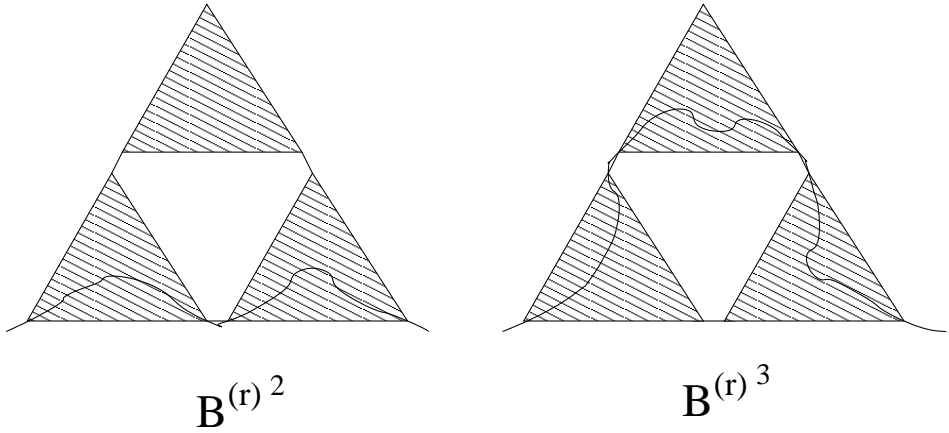}}
\caption{Two possible ways of getting a polymer configuration of type $B$
\label{3simplex3}}
\end{figure}

It is straight forward to write down the recursion equations for these
weights $A^{(r+1)}, B^{(r+1)},$ $ C^{(r+1)}$ and $D^{(r+1)}$ in terms of the
weights at order $r$. For example, Fig. \ref{3simplex3} shows the only two
possible ways one can construct a polymer configuration of type $B$.

\begin{equation}
B^{(r+1)} = {B^{(r)}}^2 + {B^{(r)}}^3.
\label{eq13}
\end{equation}

Similarly, we get

\begin{equation}
\left(\begin{array}{ll}  A^{(r+1)}\\ C^{(r+1)}\\ \end{array}\right) =
\left(\begin{array}{ll}  1+2B^{(r)}+2{B^{(r)}}^2~~~2 {B^{(r)}}^2\\~~~~~~~~
{B^{(r)}}^2~~~~~~~~~~~~~ 3 {B^{(r)}}^2 \\ \end{array}\right)  
\left(\begin{array}{ll}   A^{(r)}\\ C^{(r)}\\
\end{array}\right) 
\end{equation} 
and
\begin{eqnarray}
D^{(r+1)} &=&  {A^{(r)}}^2 +2 {A^{(r)}}^2 B^{(r)} + 4 A^{(r)} B^{(r)}  C^{(r)}
 \nonumber \\
& &+ 6 B^{(r)} {C^{(r)}}^2+ D^{(r)} ( 2 B^{(r)} + 3 {B^{(r)}}^2)
\end{eqnarray}

We note that the equation (\ref{eq13}) has one nontrivial fixed point $B =B^*=
\frac{\sqrt{5} -1}{2}$. For starting value $x < B^*$, the recursions give
$B^{(r)} \rightarrow 0$ as $r$ tends to infinity, but for $x > B^*$,
$B^{(r)}$ increases to infinity for large $r$. This gives the connectivity 
constant $\mu = 1/B^* = (\sqrt{5}+1)/2$, the golden mean. Linearizing the 
recursion equation about this nontrivial fixed point, we see that deviations 
from the fixed point value increases as 
$B^{(r+1)} - B^* \approx \lambda_1 ( B^{(r)} - B^*)$, with
$\lambda_1 = 2 + \mu^{-2}$. This then implies \cite{dd78} that 
$\nu_a = \log 2/  \log \lambda_1 \approx 0.79862$.

The critical exponent $\gamma$ is determined in terms of the larger
eigenvalue $\lambda_2$ of the $2 \times 2$ matrix in Eq. 15 evaluated 
at the non-trivial fixed point \cite{dd78}, $(A^{*}, 
B^{*},C^{*}) = (0,1/\mu,0)$. We get
\begin{equation}
\gamma_a= \frac{\log ( \lambda_2^2/3)}{\log \lambda_1} \approx 1.37522.
\end{equation}

\begin{figure}
\epsfxsize=.7\hsize\centerline{\epsfbox{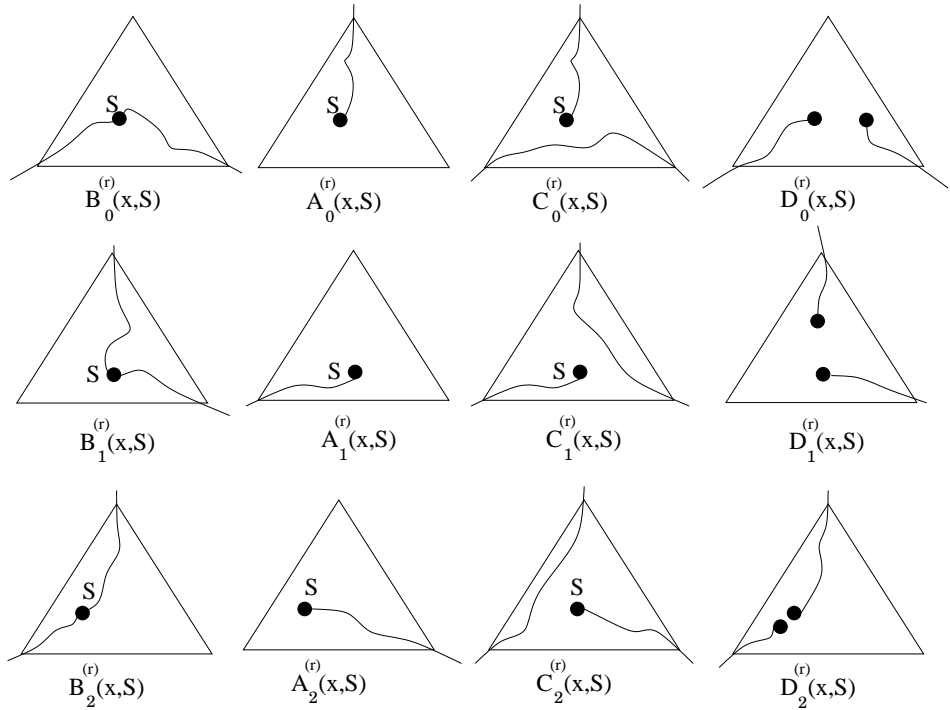}}
\caption{Definition of weights for rooted graphs. In case of $D$, $S$ can be 
any of the two end-points of the walk (shown by filled circles in the figure).
\label{3simplex4}}
\end{figure}

The recursion equations for rooted SAP's and SAW's are constructed
similarly.  We define $B_s^{(r)}(S)$ for $s=0, 1$ and $2$ as the sum over
walks on the $r$ th order triangle that go through two corners of the
triangle, visit the site $S$, avoiding the top, left and right corners for
$s=0, 1$ and $2$ respectively (Fig \ref{3simplex4}). Since $B^{(r)}_s(S)$
depends on $S$ only through $[S]_{r-1}$, we shall  write  $B^{(r)}_s(S) =
B^{(r)}_s([S]_{r-1})$. For any given $r$ and $S$, these are finite degree
polynomials in $x$. Similarly, we have to define three functions 
$A_s^{(r)}([S]_{r-1})$ and $C_s^{(r)}([S]_{r-1})$ for $s=0,1,2$ instead of the 
single variables $A$, $C$ and $D$ for the unrooted problem, as the root 
breaks the symmetry between the corner sites of the $r$-th order triangle 
(Fig. \ref{3simplex4}).

A site in the $(r+1)$-th order triangle is characterized by a string of $r$
characters. Hence, a site characterized by string $S$ at the $r$-th stage
will be characterized by one of the strings $0S$, $1S$ or $2S$ at $(r+1)$ th
stage.

\begin{figure}
\epsfxsize=.7\hsize\centerline{\epsfbox{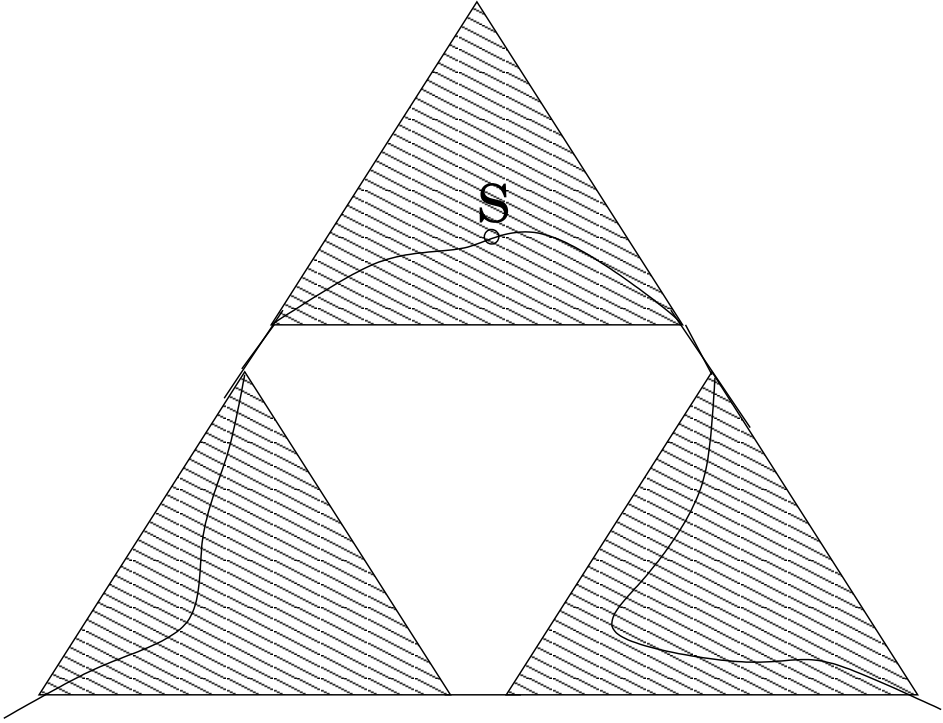}}
\caption{The only configurations contributing to $B_0^{(r+1)}(x,0S)$. Here
each sub-triangle is a $r^{th}$ order $3$-simplex.
\label{3simplex5}}
\end{figure}

We can now write the recursions similar to Eq. \ref{eq13} for
$B_s^{(r+1)}([S]_r)$, for $s= 0,1$ and $2$ in terms of weights of $r$th
order graphs. For example, one can construct  $B_0^{(r+1)}(0S)$ in just one
way as shown in Fig. \ref{3simplex5}.  The recursion relation for
$B_0^{(r+1)}([S]_r)$ will be

\begin{equation}
B_0^{(r+1)}(0[S]_{r-1}) = (B^{(r)})^2 B_0^{(r)}([S]_{r-1})
\label{eq14}
\end{equation}

Clearly, all the recursion relations for rooted polygons are   linear in the
rooted restricted partition functions of lower order  $B^{(r)}_s([S]_{r-1})$
, and can be written in the matrix form. With $[S]_r=s_r [S]_{r-1}$, we have

\begin{equation}
\left(\begin{array}{ll}  B^{(r+1)}_0([S]_{r})\\
B^{(r+1)}_1([S]_{r})\\ B^{(r+1)}_2([S]_{r})\\ \end{array}\right) =
{\cal{M}}_{s_r}  \left(\begin{array}{ll}   B^{(r)}_0([S]_{r-1})\\
B^{(r)}_1([S]_{r-1})\\ B^{(r)}_2([S]_{r-1})\\
\end{array}\right) 
\label{eq15}
\end{equation} 

where

\begin{eqnarray*} 
{\cal{M}}_0 = \left(\begin{array}{lll} B^2 ~~0 ~~0 \\ ~0 ~~B ~~B^2\\ ~0
~~B^2 ~B\\
\end{array}\right);~~{\cal{M}}_1 = \left(\begin{array}{lll} ~B ~~0 ~B^2 \\
~~0 ~B^2 ~0\\ ~B^2 ~0 ~~B\\ \end{array}\right);
\end{eqnarray*}

\begin{equation}
{\cal{M}}_2 = \left(\begin{array}{lll} B ~ B^2 ~0 \\ B^2 ~ B ~0 \\ ~0 ~~0 ~
B^2\\
\end{array}\right).
\label{eq16} 
\end{equation} 

Here we have suppressed the superscript $(r)$ of $B$ in the matrices 
${\cal M}_0$, ${\cal M}_1$ and ${\cal M}_2$. The generating function 
for rooted polygons, rooted at a site $S$ is given by

\begin{equation} 
P(x,S) = \sum_{r=1}^{\infty}B^{(r)}_{s_{r}} ( [S]_{r-1}) {B^{(r)}}^2
\label{eq17}
\end{equation}

Similarly, we can write the recursion equations for $A_s^{(r)}([S]_{r-1})$
and $C_s^{(r)}([S]_{r-1})$ and $D_s^{(r)}([S]_{r-1})$ defined analogous to
$B_s^{(r)}([S]_{r-1})$, with $s = 0,1, $ or $2$.  The $A's$ and $C's$ depend
on each other and the recursion relations for them  are

\begin{equation}
\left(\begin{array}{ll}  A^{(r+1)}_0([S]_r)\\ A^{(r+1)}_1([S]_{r})\\
 A^{(r+1)}_2([S]_{r})\\C^{(r+1)}_0([S]_{r})\\C^{(r+1)}_1([S]_{r})\\C^{(r+1)}_2
([S]_{r})
\end{array}\right) =  {\cal{L}}_{s_r} \left(\begin{array}{ll}  
A^{(r)}_0([S]_{r-1})\\ A^{(r)}_1([S]_{r-1})\\
 A^{(r)}_2([S]_{r-1})\\C^{(r)}_0([S]_{r-1})\\C^{(r)}_1([S]_{r-1})\\
C^{(r)}_2([S]_{r-1})\end{array}\right)
\label{eq18}
\end{equation} 

where 

\begin{eqnarray*} {\cal{L}}_0 =
\left(\begin{array}{llllll} 1 & 0 & 0 & 0 & B^2 & B^2\\ 0 & B & B^2 & 0 & 0
& 0\\ 0 & B^2 & B & 0 & 0 & 0\\B^2 & 0 & 0 & B^2 & 0 & 0 \\ 0 & 0 & 0 & 0 &
B^2 & 0\\0 & 0& 0 & 0 & 0 & B^2\\
\end{array}\right);
\end{eqnarray*}

\begin{eqnarray*} {\cal{L}}_1 =
\left(\begin{array}{llllll} B & 0 & B^2 & 0 & 0 & 0\\ 0 & 1 & 0 & B^2 & 0 &
B^2\\ B^2 & 0 & B & 0 & 0 & 0\\0 & 0 & 0 & B^2 & 0 & 0 \\ 0 & B^2 & 0 & 0 &
B^2 & 0\\0 & 0& 0 & 0 & 0 & B^2\\
\end{array}\right);
\end{eqnarray*}

\begin{equation} {\cal{L}}_2 =
\left(\begin{array}{llllll} B & B^2 & 0 & 0 & 0 & 0\\ B^2 & B & 0 & 0 & 0 &
0\\ 0 & 0 & 1 & B^2 & B^2 & 0\\0 & 0 & 0 & B^2 & 0 & 0 \\ 0 & 0 & 0 & 0 &
B^2 & 0\\0 & 0& B^2 & 0 & 0 & B^2
\end{array}\right)
\label{eq19}
\end{equation}

Here again we have suppressed the superscript $(r)$ on the $B$'s. One can
now write similar recursion for $D^{(r)}([S]_r)$ also. [Like the annealed 
case \cite{dd78}, these variables are not needed for the determination of
critical exponents, though they are needed to determine  $W_n(S)$.] Here we 
write down the recursions for $D_0^{(r+1)}([S]_r)$. similar relations will 
hold for $D_1^{(r+1)}([S]_r)$ and $D_2^{(r+1)}([S]_r)$.

\begin{eqnarray}
D_0^{(r+1)}(0[S]_{r-1}) &=& (AB+2BC) (A_1+A_2)+B^2D_0\nonumber \\
D_0^{(r+1)}(1[S]_{r-1}) &=& (A+AB)A_1 +(AB+2BC)(C_1+C_2) \nonumber \\
 &+& D_0B+D_2B^2+2BCC_0 \nonumber\\
D_0^{(r+1)}(2[S]_{r-1}) &=& (A+AB)A_2 +(AB+2BC)(C_1+C_2) \nonumber \\
 &+&+D_0B+D_1B^2+ 2BCC_0
\end{eqnarray}  
where we have suppressed the $([S]_{r-1})$ dependence, and the superscripts 
${(r)}$ in the terms on the right hand side.

We can also express the open walks generating function $W(x,S)$ in 
terms of these restricted partition functions. It is easy to show that

\begin{eqnarray}
W(x;S)&=&\sum_{r=1}^{m} \left[\sum_{s^{'} \neq 
s_r}A_{s^{'}}([S]_{r-1}) 
(A+AB) +D_{s_r}([S]_{r-1}) B^2 \right] \nonumber \\
&+& A A_{s_m}([S]_{m-1}]) + {\mathcal O} (x^{2^{m-1}})
\end{eqnarray}

In the limit of large $m$, the last term can be dropped, and we get

\begin{eqnarray}
W(x;S)=\sum_{r=1}^{r=\infty} \left[\sum_{s^{'} \neq s_r}A_{s^{'}}([S]_{r-1}) 
(A+AB) +D_{s_r}([S]_{r-1}) B^2 \right]
\end{eqnarray}

For order $r$, there are $3^{r-1}$ different choices for $[S]_{r-1}$, and
$3$ choices of $s$ each of the $A_s, B_s, C_s$ and $D_s$, a total of
$4.3^{r}$ polynomials. The generating function for walks rooted at one end $
W(x,S)$ can be easily written in terms of these generating functions.   To
determine $P(x,S)$ and $W(x,S)$ to given order $n$ in $x$, we need to
determine these only upto a finite order $r$. Thus, we have an efficient
algorithm to explicitly determine these polynomials to any given order, and
also to calculate the different averages over different sites. The explicit 
enumeration of rooted walks on fractals was studied by Reis \cite{reis} and 
Ordemann et. at \cite{ordemann}, but the sizes they could reach were much 
smaller. We have obtained exact values of SAWs upto length $128$ and SAPs 
upto $768$ for all sites of the fractal, using our recursion equations 
(see Table.1)

\section{Determination of the growth constant}

We note that the recursion equation of $B^{(r)}$ (Eq. 14) does not involve
the rooted variables $B_s^{(r)}(S)$, but not vice versa. If we start with  a
value $x < B^* = 1/\mu$, then for large $r$, $B^{(r)}$ tends to zero. This
would make the rooted variables $ B_s^{(r)}(S)$ also tend to zero for large
$r$. However, if $x  > B^*$, then $B^{(r)}$ will diverge for large $r$,
and this would make $B_s^{(r)}(S)$ also diverge for all $S$. Thus we
conclude that for all $S$, $P(x,S)$ and $W(x,S)$ converge if $x < 1/\mu$,
and diverge if $x > 1/\mu$. Hence it follows that the critical growth
constant is $\mu$, for all $S$.

Consider now polygons or walks of finite length $n$, with $n$ large. We
think of $x$ as the fugacity variable for each step. Then large $n$
corresponds to initial value $x$ very near $1/\mu$, say $x = 1/\mu -
\delta$ where $\delta $ scales as $1/n$. Under renormalization, the value
of $ B^* - B^{(r)}$ increases, and there is a value $r_0$ such that for  $r
- r_0 \gg 1, B^{(r)} \approx 0$, and for $r_0-r \gg 1, B^{(r)}  \approx
B^*$. Then, the recursion equations \ref{eq15} and \ref{eq18} imply that the 
rooted partition functions $B_s^{(r)}(S)$ and  $C_s^{(r)}(S)$ also become very
small for $r >r_0$, while $A_s^{(r)}$  tends to a finite value as $r$ tends
to infinity. The value of $r_0$  increases as $\delta$ is decreased as $r_0
\approx \frac {\log  (1/\delta)}{\log \lambda_1}$, and the diameter of
polymer increases as  $2^{r_0} \sim \delta^{-\nu_a}$.  As $\delta \sim 1/n$,
this implies  that the average size of  polymer increases as $n^{\nu}$,
independent of $S$, and hence $\nu_q =  \nu_a$.

An alternative proof of the assertion that the growth constant $\mu$ is  the
same for all sites is provided by setting up upper and lower bounds  for
$P_n(S)$ and $W_n(S)$, which have the same exponential growth.

We first obtain an upper bound on $P_n(S)$. If we ignore the constraint
that the walk has to pass through $S$, we get an upper bound on the number
of such walks. For example, for walks contributing to  $B^{(r)}(S)$ we can
write $B^{(r)}_{\sigma}(x, S) \le  B^{(r)}(x)$ where $\sigma=0,1,2$ and 
the inequality between polynomials is understood to imply inequality for the 
coefficient of each power of $x$. This implies that for all sites $S$,
\begin{equation} 
P(x,S) \leq \sum_{r=1}^{\infty} {B^{(r)}}^3
\label{eq20}
\end{equation} 
If we write the function giving the upper bound in right-hand-side as  $
U(x)$, then $ U(x)$ satisfies the equation
\begin{equation} 
U(x) =x^3 +  U(x^2 + x^3)
\label{eq21}
\end{equation} 

This functional equation again has the fixed point at $x^* =1/\mu$,  and
linear analysis near the fixed point shows that $U(x)$ diverges as $- \log
(1-x \mu)$ as  $x$ tends to $1/\mu$ from below. This implies that the
coefficient of $x^n$ in the Taylor expansion of $U(x)$ varies as $\mu^n/n$
for large $n$. Thus,
\begin{equation}
P_n(S) \leq K \mu^n/n, {\rm ~~for ~all } ~n, {\rm ~and ~all } ~S,
\label{eq24}
\end{equation} 
where $K$ is some constant.

We now obtain a lower bound for $P_n(S)$. Let $S_0$ be the site whose  label
is a string with all digits $0$. This is the  the topmost site of  the
triangular graph.  For such a site $B^{(r)}_0(x;S_0)$ satisfies the
following  recursion

\begin{equation}
B^{(r)}_0(S_0) = (B^{(r-1)})^2 B^{(r-1)}_0(S_0)
\label{eq25}
\end{equation}

Clearly, for all sites $S$, we have $B^{(1)}_0 (S) \ge
B^{(1)}_0(S_0)$. Then, using Eq.(\ref{eq14}) and (\ref{eq15}), by 
mathematical induction, we see that for all $r, s$ and  $S$
\begin{equation}
B^{(r)}_s (S)  \ge B^{(r)}_0(S_0), {\rm ~ for } ~~s=0,1,2.
\end{equation}

This implies for any site $S$
\begin{equation} 
P(x,S) \geq P(x,S_0) = \sum_{r=1}^{\infty} {B^{(r)}}^2 B^{(r)}_0(S_0)
\label{eq26}
\end{equation} 

Clearly, the lower bound is actually attained for $S = S_0$. If we take
$P(x,S_0) = x L(x)$, then $L(x)$ satisfies the following  equation

\begin{equation}
L(x) = x^2+ x^2 L(x^2+x^3)
\label{eq27}
\end{equation}

Assuming that $ L(x)$ near $x = 1/\mu$ has a singular expansion of the form
$ L(1/\mu -\delta) = L(1/\mu) - K \delta^{b}$, where $K$ is some constant,
we get

\begin{equation}
b = 2 \log \mu/ \log ( 2 + \mu^{-2})=0.92717
\label{eq29}
\end{equation}
and hence
\begin{equation}
P_n(S) \geq K_1 \mu^n n^{-b-1}
\label{eq30}
\end{equation}
where $K_1$ is a constant. Hence for any $S$, we have proved

\begin{equation}
K_1 \mu^n n^{-b-1} \leq P_n(S) \leq K \mu^n/n
\label{eq31}
\end{equation}

Thus, in the limit of large $n$, for all sites $S$, we must have

\begin{equation}
\lim_{n\rightarrow \infty} \frac{\mbox{log}P_n(S)}{n} = \mu~~~
\label{eq32}
\end{equation}
 We also get the nontrivial bounds 
\begin{equation}
1-b \leq \alpha_q \leq 1
\end{equation}

Similarly it is easy to prove upper and lower bounds for open walks. A
simple lower bound is provided by the inequality $W_n(S) \geq  P_n(S)$, for
all $S$.  An upper bound to $\langle \log W_n(S)\rangle$ is  provided  by
$\log \langle W_n(S) \rangle$. But the latter is known to  vary as $ n \log
\mu + (\gamma_a -1) \log n$ for large $n$. Hence we get
\begin{equation}
\lim_{n\rightarrow \infty} \frac{1}{n} \langle \log W_n(S) \rangle  = \mu.
\label{eq33}
\end{equation}

\section{variation of $P_n(S)$ and $W_n(S)$  with $S$ and $n$ }

We use the recursion equations (19-25) to calculate the values of $P_n(S)$ 
and $W_n(S)$
for different choices of the root $S$. For an SAP of order  $r$, the minimum
perimeter is  $3 \times 2^{r-2}$. We could study these recursions upto
$r=9$ and hence took into account all  self avoiding  polygons upto sizes
$n=3 \times 2^8 = 768$. A $9$-th order triangle  has $3^8=6561$  vertices. we
calculated the different polynomials $A_s(S), B_s(S), C_s(S)$ and  $D_s(S)$
for  $s= 0,1,2$ and $3^8$ possible values of $S$, keeping all terms up to
order $x^{768}$ in each polynomial.   The coefficients for  large order
become rather large. For example,   the number of polygons of  size $768$
for the topmost site of  the fractal graph is $\approx  2.5 \times
10^{154}$. We used the  symbol manipulation software Mathematica
\cite{mathematica}, which allows  one to work with integers of arbitrary
size. These then are used to calculate various averages.

\begin{figure}
\epsfxsize=1.0\hsize\centerline{\epsfbox{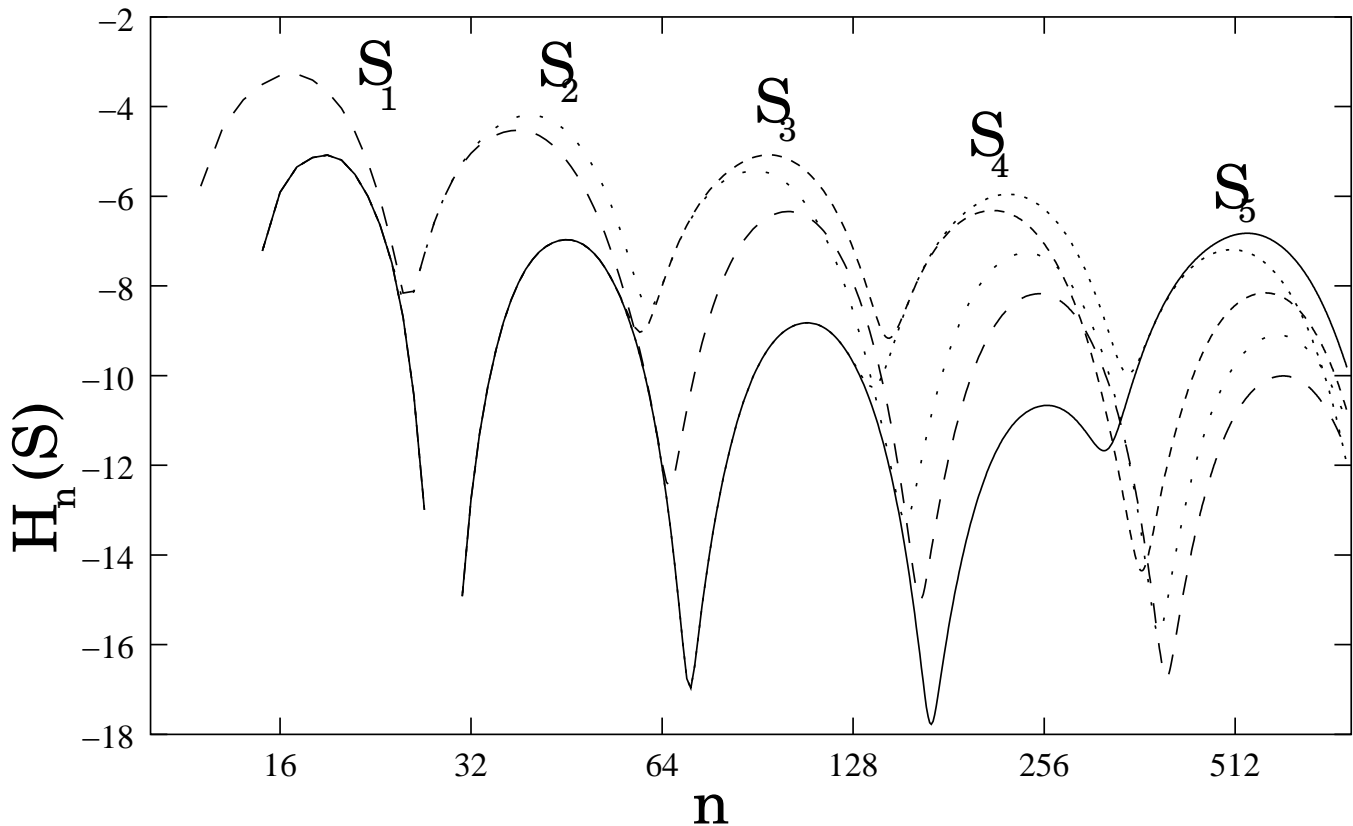}}
\caption{Figure shows $H_n(S) =logP_n(S) \mu^{-n}$ plotted as a function of 
$n$ for five different values of $S$. These $S$ are chosen near the 
constriction points and hence they maximize $P_n(S)$ for a given $n$. 
Here $S_1 = 00000211$,$S_2=00002111$,$S_3=00021111$,$S_4=00211111$ and 
$S_5=02111111$. Each of these values of $S$ attains the maximum for 
some range of $n$. The points, defined only for integer $n$, have been 
joined by straight-line segments as an aid to the eye.}
\label{ubound}
\end{figure}

In Fig. \ref{ubound}, we have shown the variation of $P_n(S) \mu^{-n}$  with
$n$ for some selected values of $S$. We see clearly roughly log-periodic
variation in these numbers. We have also plotted the upper and lower bounds
on $P_n(S)$ derived [Eq.(\ref{eq20} and \ref{eq26})]. As was argued there, 
there is a site $S_0$ which saturates the lower bound. However, there is 
no single site that saturates the upper bound. Most of the polygons having 
a given value of perimeter $n$, have a particular order $r$, and have to 
pass through the three bonds joining the $(r-1)$-order subgraphs. The six 
sites that are at the ends of these bonds clearly maximize $P_n(S)$. The 
sites change if $n$ changes to correspond to polygons with one higher order.

\begin{figure}
\epsfxsize=1.0\hsize\centerline{\epsfbox{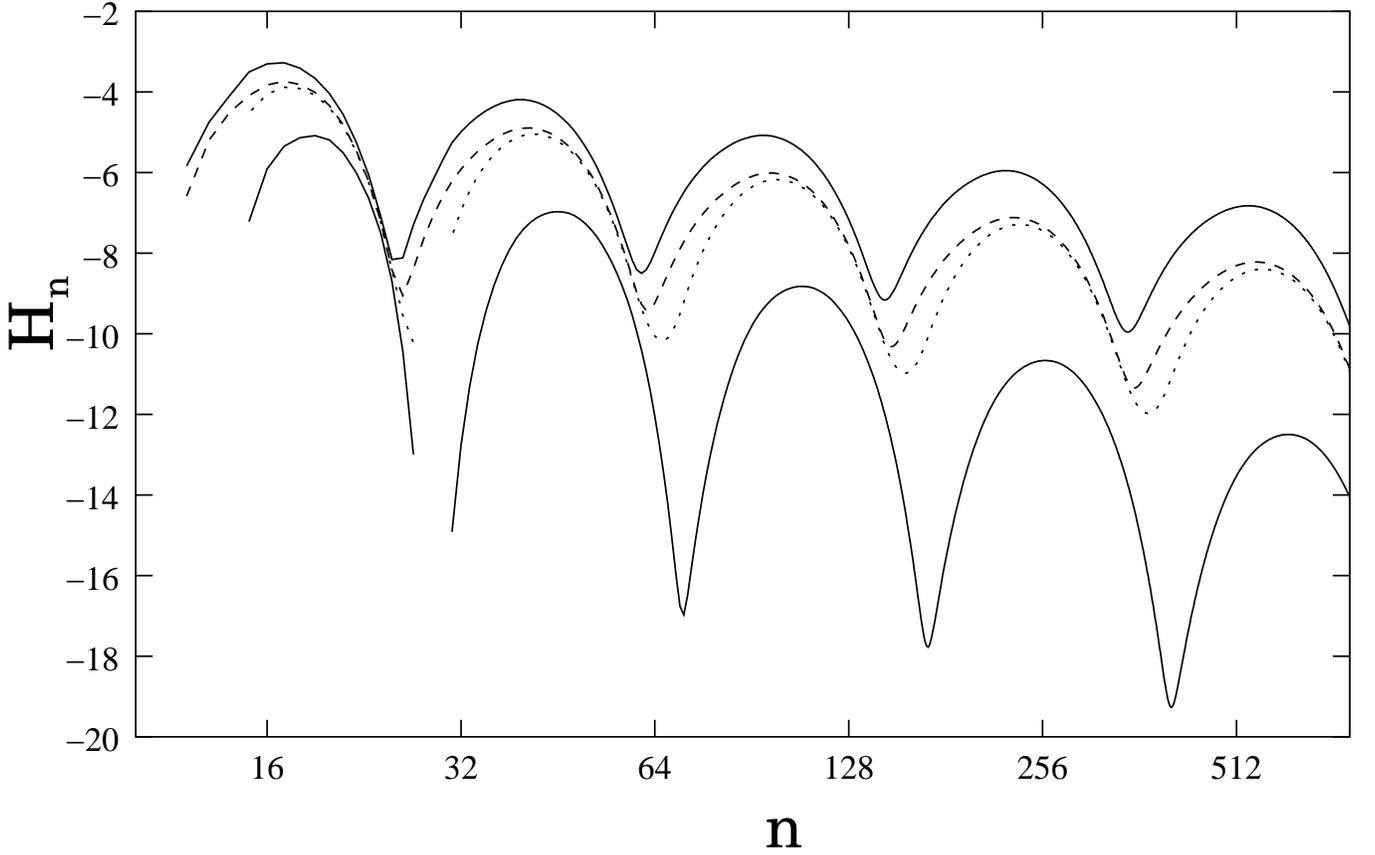}}
\caption{A plot of average value of $logP_n(S) \mu^{-n}$ ($H_n$) as a function 
of $n$. The uppermost and the lowermost curves are theoretically derived upper
and lower bounds to this number over different positions. The dashed and
dotted lines show the annealed and quenched average value respectively.
\label{quenched2}}
\end{figure}

In Fig. \ref{quenched2}, we show the variation of exactly calculated values of 
$log <P_n(S)>$ and $<log P_n(S)>$ with $n$. We also show the maximum and 
minimum values of $logP_n(S)$ attained, as a function of $n$, as $S$ takes 
all possible values. Note that though we could study sizes upto $768$ exactly, 
it is difficult to estimate $\alpha_a$ and $\alpha_q$ even to two digit 
precision from the data using standard series extrapolation techniques because 
of log-periodic oscillations.

The log-periodic variation of $P_n$ is easy to understand qualitatively 
\cite{odlyzko,logperiodic}. All SAP's of order $r$ in a given $(r+1)$th order 
triangle have to pass through the three constriction points where the 
constituting $r$-th order triangles are joined. There is a natural length 
$n \sim C \lambda_1^r$, where $C$ is some constant, for a polymer that does 
this. If the length is somewhat smaller than this value, the polymer is a bit 
stretched, and has a lower entropy. If it is a higher by a factor $1.5$ or so, 
it loses entropy as many monomers have to squeeze in the same space. If the value
of $n$ increases by a factor $\lambda_1$, then same thing happens at a
higher order triangle.  The sites which maximize $P_n(S)$ for a given $n$
are the at the constriction points of the the corresponding $r$-th order
triangles. As $\log n$ changes, these points also change. The fractional
number of points $S$ for which $P_n(S)$ attains its maximum value clearly
varies as $3^{-r}$, where $r$ is the order of a typical loop of perimeter
$n$. Using $ n \sim R^{1/\nu}$, we see that the density of points varies
as $ R^{-D} \sim n^{-D/\nu}$, where $d = {\log 3}/{ \log 2}$ is the
fractal dimension of the lattice.

\begin{figure}
\epsfxsize=1.0\hsize\centerline{\epsfbox{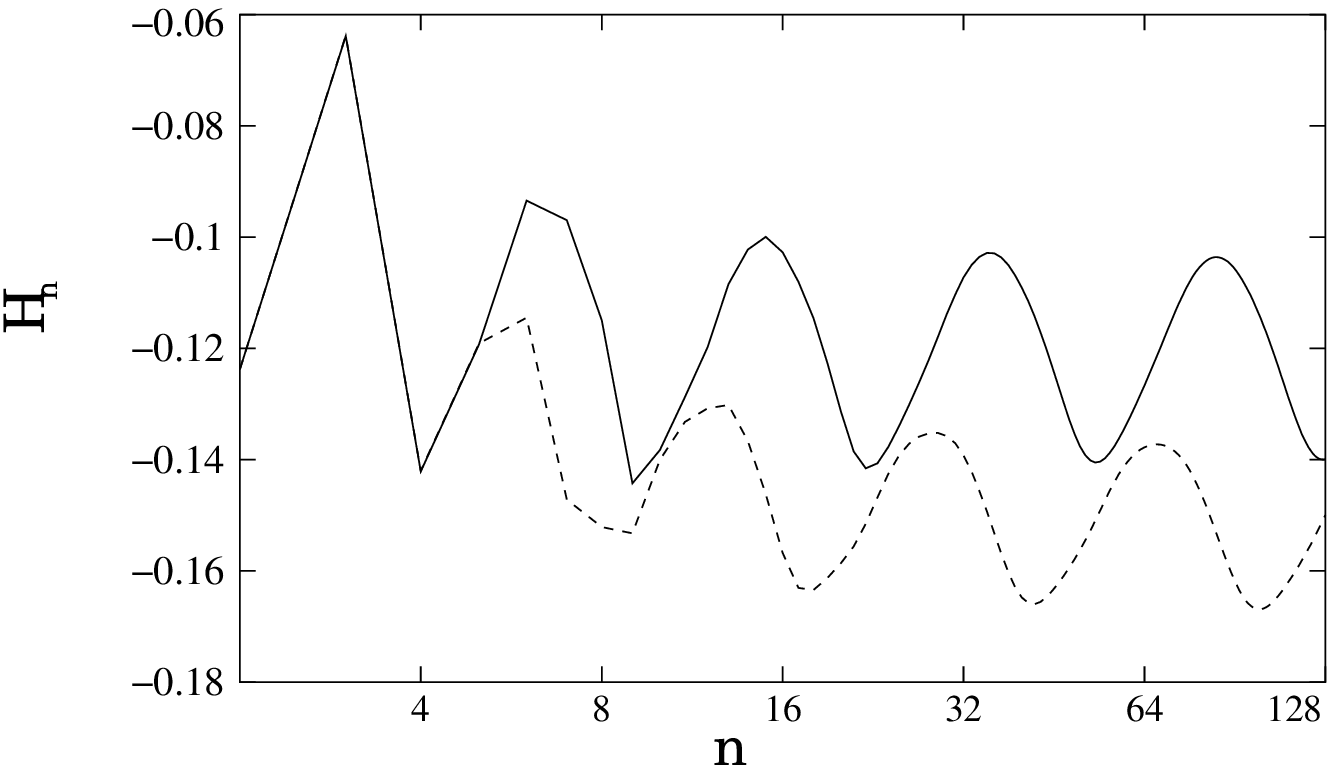}}
\caption{The figure shows plots of $H_n =log<W_n(S)>-C_n$(solid line) and 
$<logW_n(S)>-C_n$ (dotted line), where $C_n = n log \mu +(\gamma_a-1) log n$.
\label{quenwalk}}
\end{figure}

The log-periodic oscillations for open walks are much smaller in magnitude 
than in case of polygons. Here also, the number of open walks or order $r$ 
is greater if the root is near the corner point of an $r^{th}$ order triangle, 
as then it gets more space to explore and hence more entropy. We find the 
amplitude of oscillations in $H_n$ is approximately $100$ times smaller 
than for the corresponding quantity for closed polygons shown in Fig. 
\ref{ubound} and \ref{quenched2}. In Fig. \ref{quenwalk} we show the exact 
values quenched and annealed averages for walks upto size $128$. The difference 
in the two averages in the case of walks is smaller than for the polygons.

\begin{figure}
\epsfxsize=1.0\hsize\centerline{\epsfbox{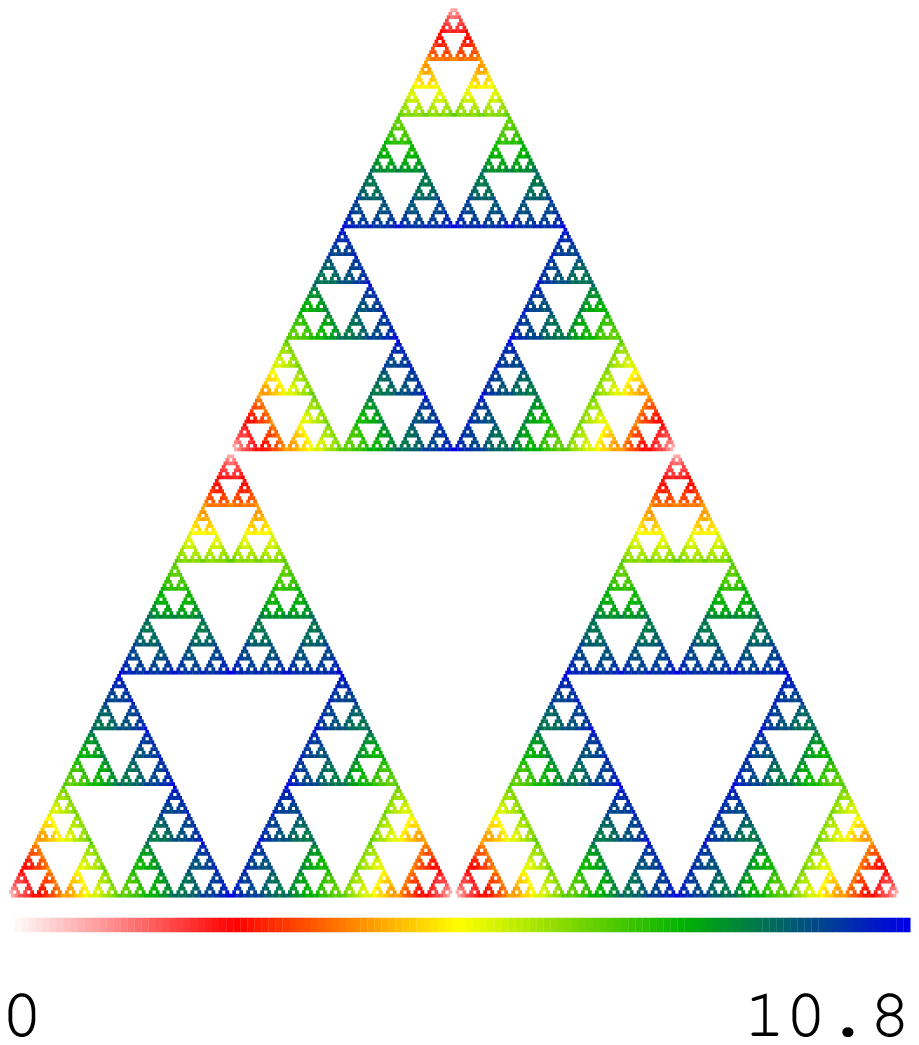}}
\caption{The plot shows the $log(P_n(S))$ for $n=400$ for all sites for
fractal  lattice upto $r=9$ generations.
\label{d400}}
\end{figure}

\begin{figure}
\epsfxsize=1.0\hsize\centerline{\epsfbox{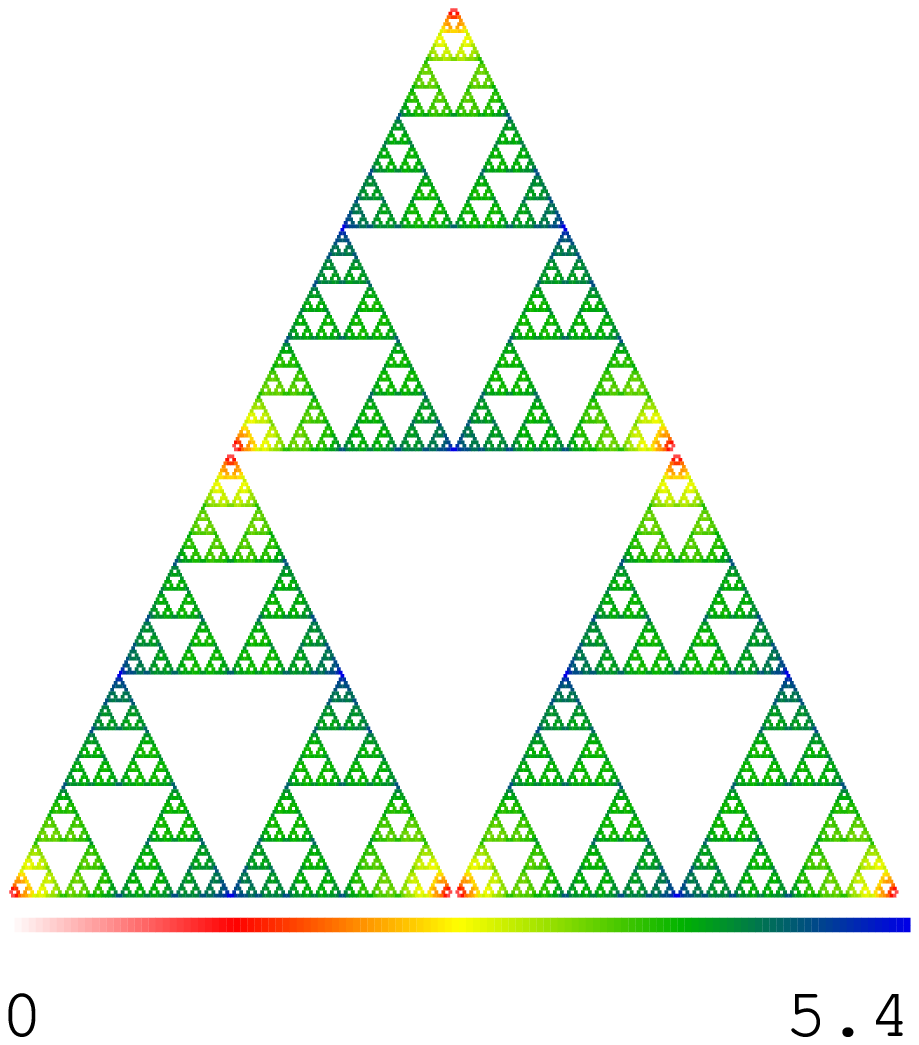}}
\caption{The plot shows the $log(P_n(S))$ for $n=600$ for all sites for
fractal  lattice upto $r=9$ generations.
\label{d600}}
\end{figure}

In Figs. \ref{d400} and \ref{d600} we have shown the density plots for SAPs
 on fractal  lattices for $n=400$ and $n=600$, The different sites are shown
 by  different colours  depending on the value of $\log P_n$ at that site.
 The  blue regions represent sites with largest values of $P_n(S)$. We
 clearly  see that depending on value of $n$ some sites are more favoured
 than others. We note that the difference between maximum and minimum values
 of  $\log P_n(S)$ is more than 10 for $n=400$, but only about half of 
this value   for $n = 600$.

\section{Calculation of exponents for quenched averages}

We note that to a very good approximation, all loops of a given perimeter
$n$ have the same order , say $r_0$, where $r_0$ is an integer  approximately
equal to $ \frac{\log n}{\log \lambda_1}$.  The contribution of these loops
to $P(x;S)$ then comes  mostly from  the single term $T_{r_0} =
B^{(r_0)}_{s_{r_0}} ( [S]_{r_0-1})  {B^{(r_0)}}^2   $ corresponding to $r =
r_0$  in  the equation (\ref{eq17}). Also, if $x  =1/\mu = B^*$, then the
contribution is $P_n(S) \mu^{-n}$, which is a slowly varying function of
$n$.  The contribution of nearby values of $n$  to the $r_0$-th term will be
comparable. The number of such terms that  contribute to the $r_0$-th term
is of order $n$. Thus we have,

\begin{equation}
P_n(S) \mu^{-n} \sim  \frac{1}{n} T_{r_0}(S)
\end{equation}

For $x= B^*$, the matrices ${\cal M}_0, {\cal M}_1, {\cal M}_2 $ become
independent of $r$, and  $T_{r_0}$ is of the form
\begin{equation}
T_{r_0} =  \langle v_1|  {\cal M}_{s_{r-1}} \ldots  {\cal M}_{s_4}~ {\cal
M}_{s_3}~ {\cal M}_{s_2}~ {\cal  M}_{s_1}| v_2\rangle
\end{equation}

Here $\langle v_1|$ and $|v_2 \rangle$ are specific $3$-dimensional  bra- 
and
ket vectors. Then from the general theory of random product of  matrices
\cite{randomproduct1,randomproduct2,randomproduct3}, it follows that the 
probability distribution  of $T_{r_0}$ tends
to a log-normal distribution for large $r_0$, and
\begin{equation}
\lim_{r_0 \rightarrow \infty} \frac {1}{r_0} \log T_{r_0} = \log \Lambda_1
\end{equation}
where $\log \Lambda_1$ is the largest Lyapunov exponent for the random 
product of matrices. This also implies that the variance of $\log P_n(S)$ will 
be proportional to $\log n$. This is much less than some positive power of $n$, 
which is the expected behaviour for the usual polymer in the random medium 
problem. For example, for the directed polymer in a random medium, the variance 
of free energy of an $n$ monomer chain varies as $n^{2/3}$. This is due to the 
fact that in the deterministic fractal case studied here, the favourable and 
unfavourable regions are very evenly distributed.

It then follows that   $\langle \log [P_n(S) \mu^{-n}] \rangle $ tends to 
$r_0 \log \Lambda_1-\log n$.  Putting $r_0 \approx (\log n)/(\log \lambda_1)$,  
we get

\begin{equation}
 \alpha_q = 1+ \log \Lambda_1  / \log \lambda_1
\end{equation}

It is straightforward to estimate $\Lambda_1$ numerically.  We start with
an arbitrary initial  3-dimensional vector  $|v_0\rangle $ of positive 
elements with sum $1$, and evolve it
randomly by the rule
\begin{equation}
|v_{j+1}\rangle = a_j  M_j |v_j\rangle
\label{ita}
\end{equation}
where $M_j$ is randomly chosen to be one of three matrices $M_0,M_1, M_2$ and 
$a_j$ is a multiplying factor chosen so that the sum of elements of new 
vector is again $1$.  The $M_j$ are

\begin{eqnarray*} 
M_0 = \left(\begin{array}{lll} {B^{*}}^2 ~~0 ~~~~0 \\ 0 ~~~~B^{*} ~~~{B^{*}}^2\\ 0~~~~{B^{*}}^2 ~B^{*}\\
\end{array}\right);
\end{eqnarray*}
\begin{eqnarray*}
M_1 = \left(\begin{array}{lll} B^{*} ~~~0~~~~{B^{*}}^2 \\0 ~~~~~{B^{*}}^2~~0\\ {B^{*}}^2 ~~0~~~~B^{*}\\ \end{array}\right);\end{eqnarray*}

\begin{equation}
M_2 = \left(\begin{array}{lll} B^{*} ~~~ {B^{*}}^2 ~~~0 \\ {B^{*}}^2 ~~ B^{*} ~~~~0 \\0 ~~~~~~0 ~~~~~{B^{*}}^2\\
\end{array}\right).
\end{equation}
where $B^{*} = (\sqrt{5}-1)/2$. 

We iterate eq.\ref{ita} many times, and  estimate $\log \Lambda_1$ by  
$\log \Lambda_1 \approx\frac{1}{j_{max}}  \sum_{j=1}^{j_{max}}  \log a_j$. 
The error in $\log \lambda_1$ decreases as $\sigma/\sqrt{j_{max}}$, where 
$\sigma$ is the rms fluctuation of $log a_j$. Values of 
$j_{max} \sim 10^8$ require less than a minute of CPU. For $j_{max} =10^9$, we 
find $\sigma \approx 0.45$ and
\begin{equation}
\log \Lambda_1 = -0.23575 \pm 0.00001
\end{equation}

Note that this corresponds to a lattice  with  $3^{j_{max}}$ sites. This gives 
$\alpha_q = 0.72837 \pm 0.00001$.

A similar calculation for the exponent $\gamma_q$ can also be done. To 
calculate the exponent $\gamma_q$, we note that again the most contribution 
to  $W(x;S)$ will come from terms with $r \approx r_0$. At $x=B^{*}$, 
${\cal L}_0$,${\cal L}_1$ and ${\cal L}_2$ become independent of $r$. If

\begin{equation}
U_{r_0} =  \langle u_1|  {\cal L}_{s_{r-1}} \ldots  {\cal L}_{s_4}~ {\cal
L}_{s_3}~ {\cal L}_{s_2}~ {\cal  L}_{s_1}| u_2\rangle
\end{equation}
where $<u_1|$ and $|u_2>$ are $6$-dimensional bra- and ket vectors and if 
$\log \Lambda_2$ is the largest Lypanouv exponent for the random product 
of matrices,

\begin{equation}
\lim_{r_0 \rightarrow \infty} \frac {1}{r_0} \log U_{r_0} =  \mbox{log}\Lambda_2
\end{equation}

Hence, $\log A_i^{(r)}(S)$ and $\log C_i^{(r)}(S)$ will increase as $r 
\log \Lambda_2$, for almost all $S$,  for $r<r_0$.
For $r>r_0$, $A_i^{(r)} \approx A_i^{(r_0)}$ and $C_i^{(r)} \approx 0$. 
Therefore it follows that the leading order contribution to  equation (29) 
will come from $r=r_0$ term. Hence $<logW_n(S) \mu^{-n}>$ tends to 
$log(\lambda_2 \Lambda_2) r_0 - log(n)$ and we get

\begin{equation}
\gamma_q = \frac{\log(\lambda_2 \Lambda_2)}{\log(\lambda_1)}
\end{equation}

We find for $j_{max}=10^9$, $\sigma \approx 1$ and $\Lambda_2 = 1.04845\pm 0.00003$. Substituting in equation above we get $\gamma_q = 1.37501\pm 0.00003$.

\begin{table}
\begin{center}
\begin{tabular}{|l|l|l|l|l|} \hline
$n$ & $log<P_n>-nlog \mu$ & $<logP_n>-nlog \mu)$ &$log<W_n>-nlog \mu$ & 
$<logW_n>-nlog \mu$ \\ \hline
3  &-1.4431783323 &  -1.4442955737  & 0.3481239636 &  0.3481239636   \\ \hline
7  &-2.5207277722 & -2.6372273881   & 0.6327712915 &  0.5825518549 \\ \hline
8  &-2.8684082046 & -2.8875491290   & 0.6648211368  &  0.6276714199  \\ \hline
9  & -4.3304492827 & -4.3328867211   &0.6797287773 &   0.6707214751  \\ \hline
15 &-4.0919348312 &-4.5179887766  &  0.9155729406  &  0.8693414702   \\ \hline
16 &-3.8277310472 &-4.0586132570  &0.9370278020  &  0.8829866367  \\ \hline
17 &-3.7374926267  & -3.8857022562 & 0.9544167885 & 0.8994169039  \\ \hline 
18 & -3.8145473883 &  -3.9148386256  & 0.9693553123  &  0.9204691093 \\ \hline
19 &-4.0133264968  & -4.0828080387 & 0.9815329752 & 0.9429045509  \\ \hline
20 & -4.3366352924  & -4.3875115144 & 0.9920263470 & 0.9648015516  \\ \hline
40 &-4.8957208753 & -5.0810665168 &1.2742813939 & 1.2185712350 \\ \hline
80 &-6.6481626291 & -7.2828262546 &1.5384807521 & 1.4965133405 \\ \hline
120 &-7.0076576021& -7.0786974211 & 1.6575701703 & 1.6393667050 \\ \hline
160 &-9.5287171328 &-10.9169115887  & & \\ \hline
320 &-9.6664681509 &-9.7339866481 & &\\ \hline
640 & -8.6892637799 &-8.7886721125 & & \\ \hline
768 &-10.8851067629 & -10.9571396026  & & \\ \hline
\end{tabular}
\vspace{0.5cm}
\caption{Values of quenched and annealed averages for SAP and SAW for some representative values of $n$. }
{\label{abcdef}}
\end{center}
\end{table}

\end{document}